\documentclass [12pt,preprint]{aastex}

\begin{document}

\title{The Chemical Evolution of Galaxies: The Stellar Mass-Metallicity 
Relation}

\author{F.D.A. Hartwick}

\affil{Department of Physics and Astronomy, \linebreak University of 
Victoria,
Victoria, BC, Canada, V8W 3P6}

In a previous paper (Hartwick 2015, H15) a model to describe the chemical 
evolution
of local dwarf galaxies was presented. Accreting gas undergoes multiple 
starbursts resulting in metallicity distributions which resemble those in 
both dwarf irregular and dwarf spheroidal galaxies. The model also predicts 
a stellar mass-metallicity (hereafter MZ) relation. The purpose of this note 
is to consider some implications of this MZ relation by comparing with the 
observations of star-forming galaxies

As discussed in H15, the logarithmic slope of the MZ relation depends on 
the assumed form of the metallicity kernel. Two kernels were considered. 
Here we use the Gaussian kernel which was introduced in the appendix to H15.
It is the most physically motivated and the predicted slope is in better 
agreement
with the more extensive observations discussed here and with the results of 
Andrews $\&$ Martini (2013). Furthermore, in order to maintain the constraint 
on the relation which holds between the mean metallicity and the variance of 
the metallicity distributions established by Leaman (2012) the mean in the 
Gaussian kernel is assumed to be zero. The convolution integrals then become 
\begin{equation}
I_{n}=C_{rs}\int_{0}^{p}z^{n-1}e^{-z^{2}/2b}dz
\end{equation}
where the dispersion b is expressed in terms of the effective yield $p_{eff}$ 
i.e. $2b=p_{eff}^{2}$ and $Log~p_{eff}$ is the metallicity at the peak of the 
metallicity distribution. The true nuclear yield is given by p and is assumed 
to be constant. $C_{rs}$ is a constant for a given redshift or epoch and is 
equal to $M_{n}(z_{\odot})/(I_{2}(z_{\odot})/p)$ where $log~z_{\odot}=log~
\overline{z}(p_{eff})=
12+log~(O/H)_{\odot}$. $M_{n}$ is the stellar mass at $z_{\odot}$ for a 
particular redshift and is determined from the observations. The values of 
$log~M_{n}/M_{\odot}$ used here in order of increasing redshift are 8.35, 9.20,
and 10.6.

The data that we will be discussing refers to the mean gas phase metallicity 
($\overline{z}_{M_{g}}$) which following the procedure in H15 is given as 
\begin{equation}
\overline{z}_{M_{g}}=(I_{2}-I_{3}/p)/(I_{1}-I_{2}/p)
\end{equation}
while the stellar mass is given by
\begin{equation}
M_{s}=I_{2}/p
\end{equation}

It is the relation between $log~\overline{z}_{M_{g}}$ and $log~M_{s}$ 
which is now compared to the observations. This relation is shown in Fig. 1 
for different values of the constant $C_{rs}$. Quenching of the starbursting 
phase at the metal rich end occurs because stars and the gas cannot be more 
metal rich than the true nuclear yield which in this case is assumed to be 
$log~(p/H)-log~(O/H)_{\odot}=0.75$. Given the nature of our model it is 
appropriate to compare the results to the observations of star-forming 
galaxies.

Zahid et al (2014) have shown from the observations of star-forming galaxies 
that {\it {at a given cosmic epoch}} the logarithmic slope of the MZ relation 
approaches $\sim 0.5$ as the metallicity decreases. The logarithmic slope of 
our model equals 0.5 at lower metallicities. The two red arcs in Fig. 1 
show the fits of their model to their observations of star-forming galaxies 
at redshifts of 0.08 and 1.55 (left to right). The constant $C_{rs}$ has
been adjusted to smoothly extend these relations to lower metallicities with 
the H15 model. The combined effect of the epoch dependence of $C_{rs}$ 
and the quenching 
of our model at the metal rich end implies that downsizing (Cowie et al 1996) 
must occur. Note that reasonable fits to the metallicity distributions in H15 
were made assuming that $C_{rs}$ is a constant. This suggests that the time 
scale for the main star forming phase is short compared to the cosmic time 
scale. 

The downsizing factor $C_{rs}$ is the same for galaxies of all metallicities at
a given epoch. This constraint may provide clues to the mechanism which 
drives the accretion e.g. gas rich mergers or colliding gas streams. 

Also shown in Fig. 1 as blue points are the observations of Onodera et al 
(2016) of star-forming 
galaxies at redshift $\sim3.3$. The line through the data is the model with 
the constant $C_{rs}$ shifted by $\delta (log M_{\ast}/M_{\odot})=1.4$ from 
the line at redshift 1.55. The green points in Fig. 1 are the observations of 
Lee et al (2006) for a sample of dwarf irregular galaxies within 5 Mpc. Here,
they are considered post-star-forming galaxies in the same sense as the dwarf 
irregular galaxies discussed in H15. Stellar ages are required in order 
to determine whether such observations belong on this plot. The prediction is 
that their major star-forming phase  
occured between redshifts 0.08 and 1.55 corresponding to look back times 
between 1.5 and 9.6 Gyr. Similarly the most massive galaxies $\sim 10^{12}M_{
\odot}$ are predicted to have undergone their main star-forming phase at 
redshift $\sim 3.3$ corresponding to a look back time of $\sim 12$ Gyr.

In summary, with the addition of one parameter ($C_{rs}$) the H15 model can 
account for the observations of the star-forming galaxies if  
the observations of the galaxies are identified with the star 
bursting behaviour of the model. This consistency with the model holds 
only if galaxy downsizing is assumed to occur.

\clearpage

\begin{figure}
\plotone{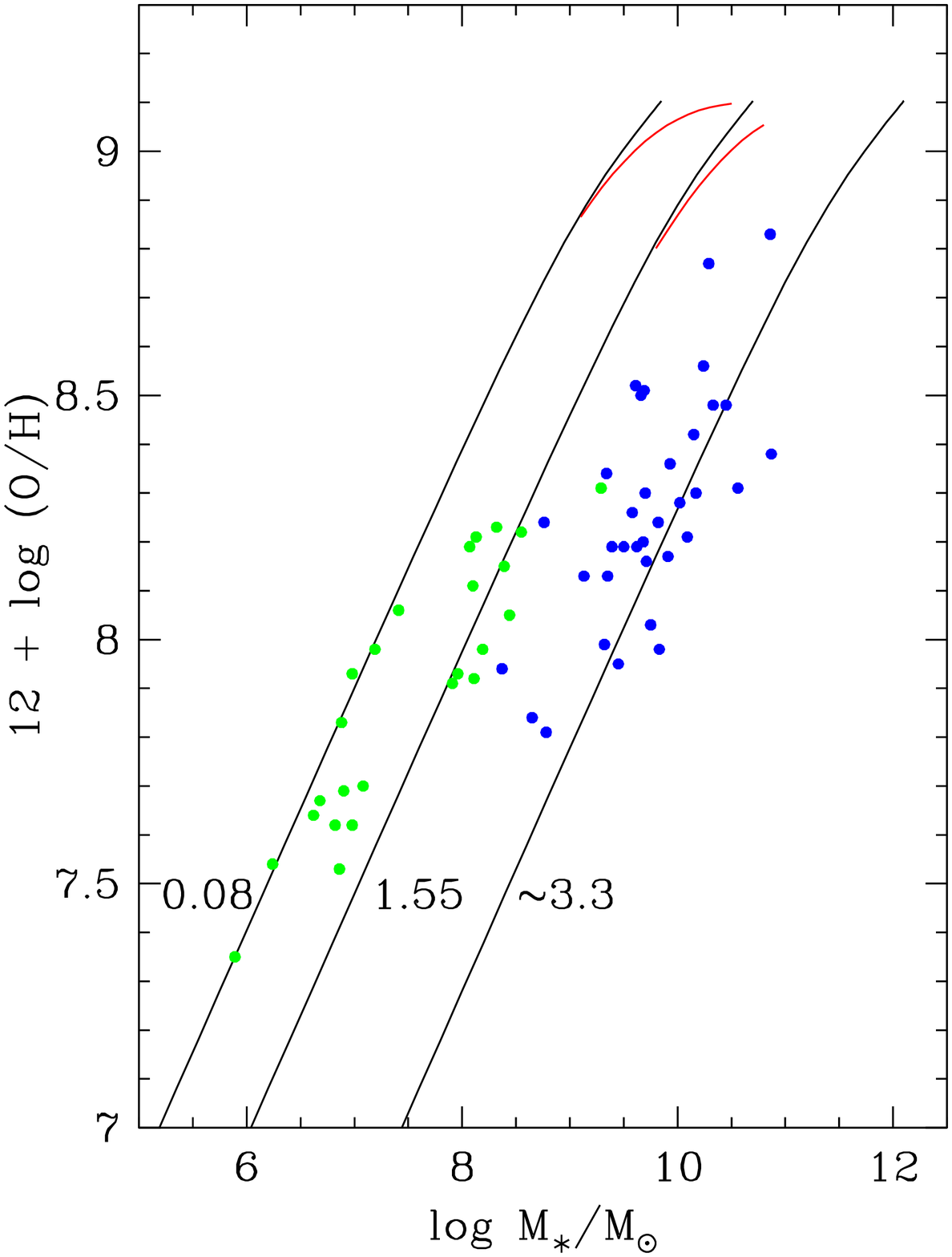}
\caption{The stellar mass-metallicity plane-the red arcs are observations from
Zahid et al (2014) for the redshifts indicated-the black lines are the model 
described above-the blue points are data for star bursting galaxies at 
redshift 3.3 from Onodera et al (2016) and the green points are for local dwarf
irregular galaxies from Lee et al (2006)}
\end{figure}
\end{document}